\documentclass[12pt]{article}

\usepackage{amsmath,amsthm,amsfonts,amssymb,amscd,amsbsy}
\usepackage{graphicx}
\usepackage{array}
\usepackage{enumerate}
\usepackage[nodayofweek]{datetime}
\usepackage[english]{babel}
\usepackage{mathtools}
\usepackage[toc,page]{appendix}
\usepackage{verbatim} 
\usepackage{amsthm} 
\usepackage{enumitem}
\usepackage{enumerate}
\usepackage{bbm} 
\usepackage[normalem]{ulem} 
\usepackage{bm}
\usepackage{authblk}

\newcommand{\be}{\begin}
\newcommand{\e}{\end}
\newcommand{\beq}{\begin{equation}}
\newcommand{\eeq}{\end{equation}}

\renewcommand{\l}{\left}
\renewcommand{\r}{\right}
\renewcommand{\d}{\mathrm{d}} 

\newcommand{\set}[1]{\mathbb{#1}}
\newcommand{\curly}[1]{\mathcal{#1}}

\newcommand{\R}{\set{R}}
\newcommand{\C}{\set{C}}

\newcommand{\eps}{\epsilon}

\newcommand{\Lam}{\Lambda}
\newcommand{\gam}{\gamma}
\newcommand{\Gam}{\Gamma}

\newcommand{\de}{\delta}

\newcommand{\scp}[2]{\langle#1,#2\rangle}
\newcommand{\scpp}[2]{\l\langle#1,#2\r\rangle}

\newcommand{\ket}[1]{|#1\rangle}

\newcommand{\ketbra}[2]{ |#1 \rangle \langle #2|}

\theoremstyle{definition}

\theoremstyle{remark}

\begin{document}

\title{Gaplessness is not generic for translation-invariant spin chains}
\date{July 15, 2019}

\author{Marius Lemm}
\affil{\textit{\small{Department of Mathematics, Harvard University, 1 Oxford Street, Cambridge, MA 02138, USA}}}

\maketitle

\begin{abstract}
The existence of a spectral gap above the ground state has far-reaching consequences for the low-energy physics of a quantum many-body system. A recent work of Movassagh [R.~Movassagh, PRL \textbf{119} (2017), 220504] shows that a spatially random local quantum Hamiltonian is generically gapless. Here we observe that a gap is more common for translation-invariant quantum spin chains, more specifically, that these are gapped with a positive probability if the interaction is of small rank. This is in line with a previous analysis of the spin-$1/2$ case by Bravyi and Gosset. The Hamiltonians are constructed by selecting a single projection of sufficiently small rank at random, and then translating it across the entire chain. By the rank assumption, the resulting Hamiltonians are automatically frustration-free and this fact plays a key role in our analysis.
\end{abstract}

\section{Introduction}
The spectral gap problem for a quantum many-body system asks the question whether its Hamiltonian operator is gapped or gapless. (The Hamiltonian is gapped, if the difference between its two lowest energy levels does not go to zero in the thermodynamic limit. If the difference goes to zero, it is gapless.)
  
The answer to the spectral gap problem has profound consequences for the low-energy physics of the many-body system. Indeed, it is known that ground states of gapped Hamiltonians display exponential decay of correlations \cite{HK06,NS06} and satisfy various notions of finite complexity in one dimension, including the famous area law for the entanglement entropy \cite{AKLV13,ALVV18,H07,LVV15}. Moreover, the occurrence of a quantum phase transition as a system parameter is varied is accompanied by a closing of the spectral gap. Hence, the spectral gap problem is intimately related to the classification of quantum phases, as has been clarified by Hastings' spectral flow method \cite{BMNS12,BHM10,H04}, also called quasi-adiabatic evolution. To summarize, many-body systems with a spectral gap are under significantly better theoretical control than their gapless counterparts. One practical consequence of this fact is that gapped systems furnish more reliable candidates for realizing quantum computation.

Given that the existence of a spectral gap is so consequential, it is unsurprising that it is in general difficult to prove rigorously. This is highlighted by the fact that Haldane's famous conjecture which predicts a spectral gap for the one-dimensio\-nal integer-spin Heisenberg antiferromagnet \cite{H83a,H83b} remains open since 1983. Significant progress towards Haldane's conjecture was achieved by Affleck, Kennedy, Lieb, and Tasaki \cite{AKLT87,AKLT88}, who in 1987 proposed an alternative isotropic and antiferromagnetic chain and showed that it has a spectral gap. The key feature of their AKLT chain that made this possible was its frustration-freeness. In frustration-free systems, the energy minimization problem defined by the Hamiltonian is inherently local, a fact that we mention because it will also play a key role in our analysis. (See \cite{Aetal} for a recent generalization of the AKLT result to two dimensions, which also relies on frustration-freeness in a fundamental way.) We also mention that one can find specific translation-invariant Hamiltonians whose spectral gap problem is equivalent to the undecidable halting problem \cite{BCLP18,CPW15}.

In this paper, we remark on a recent work of Movassagh \cite{M}, where it is shown that a completely random local Hamiltonian (meaning an independent sum of random local interactions) is gapless with probability equal to $1$, in all dimensions and under weak assumptions on the distribution of the local interaction terms. This result can be seen as surprising, since the closing of the spectral gap is usually associated with a quantum phase transition as described above. 

 Given the fact that many relevant physical systems are not spatially random, but instead translation-invariant, a natural question is then whether gaplessness is still generic in the translation-invariant class. In this paper, we prove that \emph{gaplessness is in fact no longer generic in the translation-invariant class}. We clarify that for us ``translation-invariant'' means that all the local interactions are the same, while the system may have open boundary conditions.
 
Our result applies to one-dimensional quantum spin chains which are defined by sampling a single random projection of sufficiently small rank once, and then translating it across the chain to produce a translation-invariant Hamiltonian. The result extends a much finer analysis by Bravyi and Gosset \cite{BG} of the special case where the translation-invariant chain has local spin $1/2$ (qubits) and the interaction is a rank-$1$ projector. In that case, the result of Bravyi and Gosset shows that the Hamiltonian is generically gapped. 

We use that our Hamiltonians are automatically frustration-free thanks to the small-rank assumption \cite{Metal}. This enables us to derive the spectral gap from a finite-size criterion \cite{FNW,K,LM}, which can be verified with positive probability. The proof is explicit enough to yield numerical constants and can also be used in a deterministic setting to construct gapped quantum spin chains. The method extends to trees since the small-rank assumption still implies frustration-freeness for those \cite{CM}. 

\section{Model and main results} We consider a quantum spin chain defined on $L$ sites $\{1,\ldots, L\}$ with open boundary conditions. The local Hilbert space is a qudit $\C^d$ with $d\geq 2$ and so the total Hilbert space is
\beq\label{eq:HLdefn}
\curly{H}_{\Lam_L}=\bigotimes_{j=1}^L \C^d.
\eeq
Given a fixed initial projection $P:\C^d\otimes \C^d\to \C^d\otimes \C^d$ (which will be chosen at random later), we will define the translation-invariant Hamiltonian $H_L$ by translating $P$ across the chain, i.e.,
$$
H_L=\sum_{j=1}^{L-1} P_{j,j+1}.
$$
 We will always assume that the fixed local interaction $P$ is of rank $r$, with $1\leq r\leq \max\{d-1,d^2/4\}$. This assumption ensures that $H_L$ is frustration-free \cite{Metal}, i.e., that $\ker H_L\neq \{0\}$. In other words, the ground state energy of $H_L$ is zero. The quantity of interest, the spectral gap $\gam_L$ of $H_L$, is thus equal to the smallest strictly positive eigenvalue of $H_L$.
 
We now specify the way we choose the local projection $P$ at random. Since we aim to produce random operators from random matrices, we need to fix a basis. Let us write $\{\ket{i}\}_{1\leq i\leq d}$ for the canonical basis of $\C^d$, and let $\{\ket{i\otimes j}\}_{1\leq i,j\leq d}$ be the associated basis of the tensor product $\C^d\otimes\C^d$. We order the tensor product basis lexicographically for definiteness.

\be{defn}[Random projection model]\label{defn:Prandom}
We define the local interaction $P:\C^d\otimes \C^d\to \C^d\otimes \C^d$  by
\beq\label{eq:Prandom}
P=\sum_{i=1}^r \ketbra{\phi_i}{\phi_i},
\eeq
where the states $\phi_1,\ldots,\phi_r\in \C^d\otimes \C^d$ are a random orthonormal family. The $\phi_1,\ldots,\phi_r$  family is generated as follows: Sample a random $d^2\times d^2$ orthogonal matrix $O$ from the Haar measure on the orthogonal group. Then, define the vector $\phi_i$, expressed in the canonical basis $\{\ket{i\otimes j}\}_{1\leq i,j\leq d}$, to be the $i$th column of the orthogonal matrix $O$.
\e{defn}

The Haar distribution on the orthogonal group is also called the Circular Orthogonal Ensemble (COE). The choice of an orthogonal matrix implies that the states $\phi_1,\ldots,\phi_r$ are real-valued in the canonical basis. This situation corresponds to a time-reversal symmetric Hamiltonian. We make this choice for convenience only; the methods apply equally well if the $\phi_1,\ldots,\phi_r$ are chosen as the columns of a Haar-random $d\times d$ unitary matrix (Circular Unitary Ensemble, CUE), for example.

\textit{Main results.} Our main result establishes that gaplessness is no longer generic for random projection models whose local interaction $P$ has sufficiently small rank.
 
\be{thm}[Main result]\label{thm:Prandom}
Let $P$ be as in Definition \ref{defn:Prandom} with rank $1\leq r<d$. Then the Hamiltonian $H_L$ is gapped with positive probability.
\e{thm}


In fact, we can extract numerical information from the proof of Theorem \ref{thm:Prandom}. For any $0<\eps<\frac{1}{8r}$, we have
\beq\label{eq:explicit}
\mathbb P\l(\gam_L>1-8r \eps\r)>(2d\sqrt{\pi})^{-r}\l(8\eps^2 4^{-2r}\r)^{rd^2-\binom{r+1}{2}}.
\eeq
under the probability measure given in Definition \ref{defn:Prandom}. This bound is far from optimal; we only point it out here to emphasize that the proof is hands-on.

The proof also has a deterministic version, which says that $\gam_L>1-8r\eps$ holds whenever $\|\phi_i-\ket{1\otimes (1+i)}\|<\eps$ holds for all $1\leq i\leq r<d$. 

The proof method generalizes to trees in a straightforward way. Let $k\geq2$ and let $\Gam^{(k)}_L$ be the tree which starts from a single root with $k$ children, each of which has $k$ children, etc, until the level $L$ is reached. The Hilbert space is defined by placing a qudit $\C^d$ at each site of $\Gam^{(k)}_L$. Define the Hamiltonian $H^{(k)}_L$ by placing the random interaction $P$, generated as in Definition \ref{defn:Prandom}, at each edge of $\Gam_L^{(k)}$, i.e.,
$$
H^{(k)}_L=\sum_{\substack{e=(x,y)\\ x,y\in \Gam^{(k)}_L}} P_e.
$$
From the arguments in \cite{CM}, it follows that $H^{(k)}_L$ is frustration-free if $r<d/k$. Our second main result is the following analog of Theorem \ref{thm:Prandom} on trees.

\be{thm}[Main result on trees]\label{thm:maintree}
Let $k\geq 2$ and let $P$ be as in Definition \ref{defn:Prandom} with $1\leq r<d/k$. Then the Hamiltonian $H^{(k)}_L$ is gapped with positive probability.
\e{thm}

We close the introduction with a remark about terminology.

 \be{rmk}[Boundary conditions]
 It is common terminology in the quantum spin system context to call $H_L$ defined in \eqref{eq:HLdefn} ``translation-invariant'' even though it does not commute with translations due to the open boundary conditions. The alternative terminology ``translation-invariant in the bulk'' is also sometimes used. We mention that the argument extends to periodic boundary conditions (since Theorem \ref{thm:fs} also holds for periodic boundary conditions \cite{K}), as long as one knows that $H_L$ is frustration-free. This would require extending the arguments from \cite{Metal} to periodic boundary conditions.
 \e{rmk}
 
\section{Proof of main results}

 Let us describe the proof strategy for the main result. Our main mathematical tool is a deterministic finite-size criterion which concerns two pairs of adjacent bonds. We recall that, given two projections $Q_1$ and $Q_2$, one defines $Q_1\wedge Q_2$ to be the projection onto $\mathrm{ran}(Q_1)\cap \mathrm{ran}(Q_2)$.

\be{thm}[Deterministic finite-size criterion]\label{thm:mainfs}
If we have $\|P_{1,2}P_{2,3}-P_{1,2}\wedge P_{2,3}\|<\frac{1}{2}$, then $H_L$ is gapped.
\e{thm}

We emphasize that the operators $P_{1,2}P_{2,3}$ and $P_{1,2}\wedge P_{2,3}$ act on $3$ sites only, so Theorem \ref{thm:mainfs} is indeed a finite-size criterion for gappedness. The point of Theorem \ref{thm:mainfs} is that its condition can be verified for the random projectors $P$  with positive probability. Theorem \ref{thm:mainfs} is shown by combining a Knabe-type argument for open boundary conditions \cite{K,LM} with a general bound on the anticommutator of two projections by Fannes, Nachtergaele, and Werner \cite{FNW}.

Given Theorem \ref{thm:mainfs}, the key observation is that the finite-size criterion is verified if $P_{1,2}$ and $P_{2,3}$ are almost orthogonal. This is a perturbative argument using an old theorem of von Neumann that $P_{1,2}\wedge P_{2,3}=s-\lim_{n\to\infty} (P_{1,2}P_{2,3})^n$ \cite{vN}. Given this observation it suffices to show that, for rank  $r\leq d-1$, the event that $P_{1,2}$ and $P_{2,3}$ are almost orthogonal occurs with positive probability. This is done by some elementary calculations on the sphere $S^{d^2-1}$, which we relegate to the next section. The extension to trees (proof of Theorem \ref{thm:maintree}) is postponed to the end of the paper.

We now give the details of the two proof steps: (i) deriving the deterministic finite-size criterion, and (ii), verifying it with positive probability. We denote the relevant Hamiltonian on $3$ sites by $H_{\mathrm{loc}}=P_{1,2}+P_{2,3}.$
Note that $H_{\mathrm{loc}}$ is frustration-free. We write $\gam_{\mathrm{loc}}$ for its spectral gap. The following preliminary finite-size criterion is a variant of Knabe's \cite{K} for open boundary conditions, and was observed in \cite{LM}. See also \cite{GM,L} for related criteria.

\be{thm}\cite{K,LM}\label{thm:fs}
Let $L\geq 4$.
If $\gam_{\mathrm{loc}}\geq 1$, then $\gam_L\geq 1$. If $\gam_{\mathrm{loc}}\leq 1$, then,
$$
\gam_L\geq 2\l(\gam_{\mathrm{loc}}-\frac{1}{2}\r).
$$
\e{thm}

We include the short proof of Theorem \ref{thm:fs} because it motivates the argument following it. We let 
$\{A,B\}=AB+BA$ denote the anticommutator of two matrices $A$ and $B$. 

\be{proof}[Proof of Theorem \ref{thm:fs}]
Note that the projections $P_{j,j+1}$ and $P_{k,k+1}$ commute when $|j-k|\geq 2$. This gives
\beq
\label{eq:H2}
H_L^2=H_L+\sum_{1\leq j< k\leq L-1} \{P_{j,j+1},P_{k,k+1}\}\geq H_L+Q
\eeq
where $Q:=\sum_{j=1}^{L-2}  \{P_{j,j+1},P_{j+1,j+2}\}.$ We consider the auxiliary operator
$$
\begin{aligned}
A=\sum_{j=1}^{L-2} (P_{j,j+1}+P_{j+1,j+2})^2.
\end{aligned}
$$
On the one hand, we can compute $A=2H_L-P_{1,2}-P_{L-1,L}+Q$. On the other hand, by the spectral theorem, frustration-freeness, and translation invariance, we have $(P_{j,j+1}+P_{j+1,j+2})^2\geq \gam_{\mathrm{loc}}(P_{j,j+1}+P_{j+1,j+2})$. Combining these two facts gives
$$
Q\geq  (\gam_{\mathrm{loc}}-1)\l(2H_L-P_{1,2}-P_{L-1,L}\r)
$$
If $\gam_{\mathrm{loc}}\geq 1$, then $Q\geq 0$ and \eqref{eq:H2} implies $H_L^2\geq H_L$, i.e., $\gam_L\geq 1$ by the spectral theorem and frustration-freeness. 

Let now $\gam_{\mathrm{loc}}\leq 1$. Applying the operator inequality on $Q$ to \eqref{eq:H2}, we find
$$
H_L^2\geq 2H_L\l(\gam_{\mathrm{loc}}-\frac{1}{2}\r).
$$
By the spectral theorem and frustration-freeness, this proves $\gam_L\geq 2\l(\gam_{\mathrm{loc}}-\frac{1}{2}\r)$ and hence Theorem \ref{thm:fs}.
\e{proof}

\begin{proof}[Proof of Theorem \ref{thm:mainfs}] Thanks to the preliminary finite-size criterion in Theorem \ref{thm:fs}, it suffices to prove $\gam_{\mathrm{loc}}>\frac{1}{2}$. Using that $P_{1,2}$ and $P_{2,3}$ are projections, we compute
$$
H_{\mathrm{loc}}^2=(P_{1,2}+P_{2,3})^2=H_{\mathrm{loc}}+\{P_{1,2},P_{2,3}\}.
$$
We recall Lemma 6.3 (ii) in \cite{FNW}, which says that for a pair of projections $Q_1$ and $Q_2$ on a finite-dimensional Hilbert space, one has the operator inequality $\{Q_1,Q_2\}\geq -\|Q_1Q_2-Q_1\wedge Q_2\|(Q_1+Q_2)$. This gives 
$$
H_{\mathrm{loc}}^2\geq H_{\mathrm{loc}}(1-\|P_{1,2}P_{2,3}-P_{1,2}\wedge P_{2,3}\|).
$$
By our assumption that $\|P_{1,2}P_{2,3}-P_{1,2}\wedge P_{2,3}\|<\frac{1}{2}$, we can conclude the operator inequality $H_{\mathrm{loc}}^2\geq  \frac{1}{2}H_{\mathrm{loc}}$ which is strict on $(\ker H_{\mathrm{loc}})^\perp$. By the spectral theorem and the fact that $H_{\mathrm{loc}}$ is frustration-free, this result is equivalent to $\gam_{\mathrm{loc}}>\frac{1}{2}$ and Theorem \ref{thm:mainfs} is proved.
\e{proof}

Next we prove the main result Theorem \ref{thm:Prandom} by verifying the deterministic finite-size criterion in Theorem \ref{thm:mainfs}.

 The following lemma quantifies the probability that the collection of random orthonormal vectors $\phi_1,\ldots,\phi_r$ generated by Definition \ref{defn:Prandom} lands close to a prescribed, fixed collection of orthonormal vectors $v_1,\ldots,v_r$.

\be{lm}[Probability of landing near fixed vectors]\label{lm:event}
Let $0<\eps<1/4$. Let $v_1,\ldots,v_r$ be a fixed collection of orthonormal unit vectors with real-valued coefficients in the canonical basis $\{\ket{i\otimes j}\}_{1\leq i,j\leq d}$. For the random projection model from Definition \ref{defn:Prandom}, it holds that
$$
\mathbb P\l(\max_{1\leq i\leq r}\|\phi_{i}-v_i\|<\eps\r)>(2d\sqrt{\pi})^{-r}\l(\frac{\eps/8}{4^r \sqrt{r!}}\r)^{rd^2-\binom{r+1}{2}}
$$
\e{lm}

The precise value of the positive constant on the right-hand side is not important for what follows. Notice that the event essentially says that the random vector $\phi_i$ lies in a spherical cap around $v_i$, and since a cap has positive Haar measure, it makes sense that this event will occur with positive probability. The details of the proof of Lemma \ref{lm:event} are relegated to Section \ref{sect:sphere}

We can now prove the main result by applying this lemma to a certain good collection of vectors, the $v_i$ in \eqref{eq:good} below.

\be{proof}[Proof of the main result, Theorem \ref{thm:Prandom}]
We begin by noting that our assumption $1\leq r\leq d-1$ ensures that $r\leq \max\{d-1,\frac{d^2}{4}\}$ and so $H_L$ is frustration-free thanks to \cite{Metal}. We let $0<\eps<1/4$ to be specified later and apply Lemma \ref{lm:event} with the choice
\beq\label{eq:good}
v_i=\ket{1\otimes (i+1)},\qquad \forall 1\leq i\leq r.
\eeq
Lemma \ref{lm:event} then implies that $\max_{1\leq i\leq r}\|\phi_{i}-v_i\|<\eps$ holds with positive probability, and from now on we restrict to the event where this occurs. By the triangle inequality,
\beq\label{eq:triangle}
\begin{aligned}
\|P_{1,2}P_{2,3}-P_{1,2}\wedge P_{2,3}\|
\leq& \|P_{1,2}P_{2,3}\| +\|P_{1,2}\wedge P_{2,3}\|\\
\end{aligned}
\eeq
We will now show that both of the norms on the right-hand side are bounded by a constant times $\eps$. We first estimate $\|P_{1,2}P_{2,3}\|$. We introduce the reference projection
$
\tilde P:=\sum_{i=1}^r \ketbra{v_i}{v_i}.
$
By writing $\phi_i=v_i+\varphi_i$ and using that $\|\varphi_i\|<\eps$ by assumption, we obtain that $P$ is close to the reference projection, i.e.,
\beq\label{eq:close}
\|P- \tilde P\|\leq \sum_{i=1}^r \l\|\ketbra{\phi_i}{\phi_i}-\ketbra{v_i}{v_i}\r\|
\leq 2r\eps.
\eeq
Now, the key observation is that the two reference projections are orthogonal: $\mathrm{ran}(\tilde P_{1,2})\perp \mathrm{ran}(\tilde P_{2,3}).$
Indeed, $\tilde P_{1,2}$ and $\tilde P_{2,3}$ have a common eigenbasis and orthogonal ranges in this basis. To see the latter, notice that at the middle (i.e., second) site, the states in $\mathrm{ran}(\tilde P_{1,2})$ necessarily have a label $\ket{j}$ with $j\geq 2$, while states in $\mathrm{ran}(\tilde P_{2,3})$ necessarily have the label $\ket{1}$ there. In other words, $\tilde P_{1,2}\tilde P_{2,3}=0$. Hence, \eqref{eq:close} and $\|P_{1,2}\|,\|\tilde P_{2,3}\|\leq 1$ imply
\beq\label{eq:firstpart}
\|P_{1,2}P_{2,3}\| \leq \|\tilde P_{1,2}\tilde P_{2,3}\|+4r\eps=4r\eps.
\eeq
We will choose $\eps<\frac{1}{8r}$, so that $4r\eps<\frac{1}{2}$.

It remains to estimate the other norm in \eqref{eq:triangle}, $\|P_{1,2}\wedge P_{2,3}\|$. A theorem of von Neumann \cite{vN}, page 55, says that for two projections $Q_1$ and $Q_2$ on a finite-dimensional Hilbert space, it holds that $Q_1\wedge Q_2=s-\lim_{n\to\infty}(Q_1Q_2)$. By submultiplicativity of the norm, this implies that for every vector $x$ of norm $\|x\|=1$, one has $\|(Q_1\wedge Q_2)x\|=\lim_{n\to\infty} \|(Q_1Q_2)^n x\|
\leq \liminf_{n\to\infty} \|Q_1Q_2\|^n$
and hence,  after taking the supremum over such $x$,
$
\|Q_1\wedge Q_2\|\leq \liminf_{n\to\infty} \|Q_1Q_2\|^n.
$
In our context, this fact gives
$$
\|P_{1,2}\wedge P_{2,3}\|\leq \liminf_{n\to\infty}\|P_{1,2}P_{2,3}\|^n=0,
$$
where we used \eqref{eq:firstpart} and  $4r\eps<\frac{1}{2}$ in the end. (Thus, we have in fact shown that $\mathrm{ran}(P_{1,2})\cap \mathrm{ran}(P_{2,3})=\{0\}$.) Upon returning to \eqref{eq:triangle}, we see that
$$
\|P_{1,2}P_{2,3}-P_{1,2}\wedge P_{2,3}\|<4r\eps<\frac{1}{2}.
$$
Therefore, we can apply Theorem \ref{thm:mainfs} to conclude that $H_L$ is gapped with positive probability. This proves Theorem \ref{thm:Prandom}.
\e{proof}

\section{Some geometric estimates on the sphere}
\label{sect:sphere}

In this section, we prove Lemma \ref{lm:event}. We will use the following geometrical lemma. We write $\d\mu_{n}$ for normalized Haar measure on the real $n$-dimensional unit sphere $S^n\subset \R^{n+1}$. We let $\mathbf{d}(\cdot,\cdot)$ denote the spherical distance function on $S^{d^2-1}$, i.e., arclength of great circles.

\be{lm}\label{lm:geometric}
Let $C$ be a spherical cap on $S^n$ of radius $0<\de<1/4$ in spherical distance $\mathbf{d}$. Then
$$
\mu_n(C)>\frac{1}{2\sqrt{\pi}} \frac{(\de/2)^{n}}{\sqrt{n}}.
$$
\e{lm}

\be{proof}
Notice that the height of $C$ is $h=1-\cos \de$. It is known that
$$
\mu_n(C)= \frac{1}{2}I_{2h_\de-h_\de^2}\l(\frac{n}{2},\frac{1}{2}\r)=\frac{\int_0^{2h_\de-h_\de^2} t^{\frac{n-2}{2}} (1-t)^{-1/2}\d t}{2\int_0^1 t^{\frac{n-2}{2}} (1-t)^{-1/2}\d t}
$$
where $I_{2h_\de-h_\de^2}\l(\frac{n}{2},\frac{1}{2}\r)$ is the regularized incomplete Beta function. We simplify the right-hand side by using that $2h_\de-h_\de^2\geq (\de/2)^2$ for $0<\de<1/2$, and employing known facts about the Beta and Gamma functions, in particular Gautschi's inequality, cf.\ \cite{NIST}, formula 5.6.4. We find
$$
\begin{aligned}
&\frac{\int_0^{2h_\de-h_\de^2} t^{\frac{n-2}{2}} (1-t)^{-1/2}\d t}{2\int_0^1 t^{\frac{n-2}{2}} (1-t)^{-1/2}\d t}
> \sqrt{\frac{n-1}{2}} \frac{\int_0^{(\de/2)^2} t^{\frac{n-2}{2}} \d t}{2\sqrt{\pi}}\\
&=\l(\frac{n}{2}\r)^{-1/2}\frac{(\de/2)^{n}}{2\sqrt{\pi}}
\geq \frac{1}{2\sqrt{\pi}} \frac{(\de/2)^{n}}{\sqrt{n}}.
\end{aligned}
$$
This proves Lemma \ref{lm:geometric}.
\e{proof}

We are now ready to prove the proposition.  

\be{proof}[Proof of Lemma \ref{lm:event}]
\textit{Step 1:}
Let $0<\de<1/4$. The bound $\mathbf{d}(\phi_1,v_1)<\de$ expresses precisely that $\phi_1$ lies inside of a spherical cap on $S^{d^2-1}\subset \R^{d^2}$ of radius $\de$ and center $v_1$; call it $C_1$. By rotational symmetry, the marginal measure induced on $\phi_1$ by Definition \ref{defn:Prandom} is just Haar measure on $S^{d^2-1}$. We can thus apply Lemma \ref{lm:geometric} with $n=d^2-1$ to find
\beq\label{eq:phi1prob}
\mathbb P(\mathbf{d}(\phi_1,v_1)<\de)=\mu_{d^2-1}(C_1)>\frac{1}{2\sqrt{\pi}}  \frac{(\de/2)^{d^2-1}}{d}.
\eeq

This inequality already implies the $r=1$ case in Lemma \ref{lm:event} by setting $\de=\eps$. Indeed, the inequality $\|\phi_{1}-v_1\|<\eps$ expresses that $\phi_1$ lies inside of a Euclidean ball of radius $\de$ and center $v_1$. Since the spherical distance dominates the Euclidean distance, this is implied by $\mathbf{d}(\phi_1,v_1)<\eps$, and the probability of this event is bounded from below by \eqref{eq:phi1prob} with $\de=\eps$.\\

\textit{Step 2:}
Let $r\geq 2$ and $2\leq i\leq r$. Let $\de<\frac{1}{4^{j+1}\sqrt{j!}}$. For all $1\leq j\leq r$, define $A_j$ as the event that $\mathbf{d}(\phi_j,v_j)<\de 4^j\sqrt{j!}$ holds. We claim that
\beq\label{eq:conditional}
\mathbb P(A_i\vert  A_{i-1}\cap\ldots\cap A_1)>\frac{1}{2d\sqrt{\pi}} \l(\frac{\de}{8}\r)^{d^2-i}.
\eeq
We first notice that the event $A_{i-1}\cap\ldots\cap A_1$ only concerns the random vectors $\phi_{i-1},\ldots,\phi_1$ while $A_i$ concerns only the random vector $\phi_i$. Let us now fix an outcome for $\phi_{i-1},\ldots,\phi_1$ belonging to the event $A_{i-1}\cap\ldots\cap A_1$. By rotational symmetry and orthonormality, the random vector $\phi_i$ is Haar distributed on the $(d^2-i)$-dimensional sphere consisting of those vectors in $S^{d^2-1}$ which are orthogonal to $\phi_1,\ldots,\phi_{i-1}$; call it $S_{\perp}^{d^2-i}\subset S^{d^2-1}$. 

Let $w$ denote the vector in $S_{\perp}^{d^2-i}$ that achieves the spherical distance $\mathbf{dist}(S_{\perp}^{d^2-i},v_i)$, that is, $\mathbf{d}(w,v_i)=\mathbf{dist}(S_{\perp}^{d^2-i},v_i)$. (This exists by compactness.) The idea is now that, if $\phi_i$ lands close to $w$ (which it can, since $w$ is in the allowed set  $S_{\perp}^{d^2-i}$), then it will also be close to its true target, the vector $v_i$. To formalize this idea, we define the auxiliary event $B_i$, which says that $\mathbf{d}(\phi_i,w)<\de 4^{i-1}\sqrt{i!}$. 

\be{lm}\label{lm:also}
Assume that $A_{i-1}\cap\ldots\cap A_1$ occurs. If $B_i$ occurs, then $A_i$ occurs as well. 
\e{lm}

\be{proof}[Proof of Lemma \ref{lm:also}]
Suppose that $B_i$ occurs. We need to show that $\mathbf{d}(\phi_i,v_i)<\de 2^i\sqrt{i!}$ holds. By the triangle inequality and the definition of $B_i$,
\beq\label{eq:triangle1}
\begin{aligned}
\mathbf{d}(\phi_i,v_i)
&\leq \mathbf{d}(\phi_i,w)+\mathbf{d}(w,v_i)\\
&\leq \de 4^{i-1}\sqrt{i!}+\mathbf{dist}(S_{\perp}^{d^2-i},v_i).
\end{aligned}
\eeq
It remains to bound $\mathbf{dist}(S_{\perp}^{d^2-i},v_i)$. Recall that $S_{\perp}^{d^2-i}$ is defined as the orthogonal complement of $\phi_1,\ldots,\phi_{i-1}$. Let $\sum_{j=1}^{i-1} c_j \phi_j$ be an arbitrary unit vector in $\mathrm{span}\{\phi_1,\ldots,\phi_{i-1}\}$, so $\sum_{j=1}^{i-1}|c_j|^2=1$. Then, by orthonormality of $v_1,\ldots,v_i$, 
$$
\begin{aligned}
\l|\scpp{\sum_{j=1}^{i-1} c_j \phi_j}{v_i}\r|
&=\l|\sum_{j=1}^{i-1} c_j \scp{ \phi_j-v_j}{v_i}\r|\\
&\leq \sum_{j=1}^{i-1} |c_j| \|\phi_j-v_j\|\\
&\leq \sum_{j=1}^{i-1} |c_j| \mathbf{d}(\phi_j,v_j)\\
&\leq \de 4^{i-1} \sqrt{(i-1)!}  \sum_{j=1}^{i-1} |c_j|\\
 &\leq \de 4^{i-1} \sqrt{i!}
 \end{aligned}
$$
In the second-to-last step, we used the assumption that $A_{i-1}\cap\ldots\cap A_1$ occurs, and in the last step, we applied the Cauchy-Schwarz inequality. Recall that $\mathbf{d}(x,y)=\arccos\scp{x}{y}$ for any two unit vectors $x,y$. Using that $t\mapsto \arccos(t)-\frac{\pi}{2}$ is an odd and monotonically decreasing function, as well as the estimate $\arccos(t)>\pi/2-2t$ for all $0<t<1$, we find
$$
\begin{aligned}
\l|\mathbf{d}\l(\sum_{j=1}^{i-1} c_j \phi_j,v_i\r)-\frac{\pi}{2}\r| 
&\leq \frac{\pi}{2}- \arccos(\de 4^{i-1} \sqrt{i!})\\
&\leq 2\de 4^{i-1} \sqrt{i!}
 \end{aligned}
$$
Since the coefficients $\{c_j\}$ were arbitrary, this show that $v_i$ is almost orthogonal to $\phi_1,\ldots,\phi_{i-1}$, i.e.,
$$
\mathbf{dist}(S_{\perp}^{d^2-i},v_i)<2\de 4^{i-1} \sqrt{i!}.
$$
Applying this estimate to \eqref{eq:triangle1}, we obtain $\mathbf{d}(\phi_i,v_i)<\de 4^i\sqrt{i!}$, i.e., the event $A_i$ occurs. This proves Lemma \ref{lm:also}.
\e{proof}

We continue with the proof of \eqref{eq:conditional}. On the one hand, by Lemma \ref{lm:also}, $B_i\subset A_i$ conditional upon $A_{i-1}\cap\ldots\cap A_1$ and so
$$
\mathbb P(A_i\vert  A_{i-1}\cap\ldots\cap A_1)>
\mathbb P(B_i\vert  A_{i-1}\cap\ldots\cap A_1).
$$
 On the other hand, conditional upon $A_{i-1}\cap\ldots\cap A_1$, the event $B_i$ is precisely the event that the Haar distributed vector $\phi_i$ lies in a certain spherical cap on $S_{\perp}^{d^2-i}$ of radius $\de 4^{i-1}\sqrt{i!}$; call this cap $C_i$. Hence, the conditional probability of $B_i$ can be bounded by Lemma \ref{lm:geometric} as follows
$$
\begin{aligned}
&\mathbb P(B_i\vert  A_{i-1}\cap\ldots\cap A_1)=\mu_{d^2-i}(C_i)\\
&>\frac{1}{2\sqrt{\pi}} \frac{(\de 4^{i-1} \sqrt{i!}/2)^{d^2-i}}{\sqrt{d^2-i}}\\
&>\frac{1}{2d\sqrt{\pi}} \l(\frac{\de}{8}\r)^{d^2-i}.
\end{aligned}
$$
This proves \eqref{eq:conditional} and finishes step 2.\\


\textit{Step 3:}
By the chain rule for conditional probability, we have
\beq\label{eq:chainrule}
\mathbb P(A_r\cap A_{r-1}\cap\ldots\cap A_1)
=\prod_{i=1}^r \mathbb P(A_i \vert A_{i-1}\cap\ldots\cap A_1)
\eeq
with the convention that the $i=1$ term on the right-hand side is $\mathbb P(A_1)$. Let $0<\eps<1/4$, and set $\de:=\frac{\eps}{4^r\sqrt{r!}}$. Then, it is guaranteed that $\de<\frac{1}{4^{j+1}\sqrt{j!}}$ for all $1\leq j\leq r$. Hence, we can combine \eqref{eq:phi1prob} and \eqref{eq:conditional} to conclude that
\beq\label{eq:pestimate}
\begin{aligned}
&\mathbb P(A_r\cap A_{r-1}\cap\ldots\cap A_1)\\
&>\frac{1}{2d\sqrt{\pi}}  \l(\frac{\de}{2}\r)^{d^2-1} \times \prod_{i=2}^r \frac{1}{2d\sqrt{\pi}} \l(\frac{\de}{8}\r)^{d^2-i}\\
&>(2d\sqrt{\pi})^{-r}\l(\frac{\de}{8}\r)^{rd^2-\binom{r+1}{2}}.
\end{aligned}
\eeq
Now, by definition, on the event $A_r\cap A_{r-1}\cap\ldots\cap A_1$, we have 
$$
\sup_{1\leq i\leq r} \mathbf{d}(\phi_i,v_i)< \de 4^{r} \sqrt{r!}=\eps.
$$
Finally, we recall that $\|\phi_i-v_i\|\leq \mathbf{d}(\phi_i,v_i)$ and so Lemma \ref{lm:event} now follows from the estimate \eqref{eq:pestimate} with $\de=\frac{\eps}{4^r\sqrt{r!}}$.
\e{proof}

\section{Extension to trees}
\label{sect:maintreepf}
In this section, we prove Theorem \ref{thm:maintree} by modifying the proof of Theorem \ref{thm:Prandom}. We use that $H_L^{(k)}$ is frustration-free for $r<d/k$. This follows by examining the proof on page 5 of \cite{CM}. Let us write $\gam^{(k)}_L$ for its spectral gap. 

The key step is to adapt the finite-size criterion Theorem \ref{thm:fs} to trees. Recall that $H_{\mathrm{loc}}=P_{1,2}+P_{2,3}$ acts on $3$ sites, and $\gam_{\mathrm{loc}}$ is its spectral gap.

\be{thm}[Finite-size criterion on trees]
\label{thm:fstree}
Let $L\geq 4$. If $\gam_{\mathrm{loc}}\geq 1$, then $\gam^{(k)}_L\geq 1$. If $\gam_{\mathrm{loc}}\leq 1$, then,
$$
\gam_L^{(k)} \geq 2k\l(\gam_{\mathrm{loc}}-1+\frac{1}{2k}\r).
$$
\e{thm}

Notice that the one-dimensional chain corresponds to $k=1$ and so this result reproduces Theorem \ref{thm:fs}.

\be{proof}
We follow the proof of Theorem \ref{thm:fs}. As in \eqref{eq:H2}, we have $(H^{(k)})^2\geq H^{(k)}+Q^{(k)}$, where $Q^{(k)}$ contains all the terms of the form $\{P_e,P_{e'}\}$ where the two edges $e$ and $e'$ share exactly one vertex. The main difference is that the auxiliary operator $A$ is now replaced by
$$
A^{(k)}=\sum_{e\in \Gam_L^{(k)}} \sum_{e'\sim e}\l(P_e+P_{e'}\r)^2
$$
where $e\sim e'$ means that the two edges $e$ and $e'$ share exactly one vertex and $\Gam_L^{(k)}$ are the first $L$ levels of the graph $\Gam_k$. 

On the one hand, since every edge $e$ has $k+(k-1)+1=2k$ partners $e'\sim e$, we find that
$$
A^{(k)}=2k \tilde H^{(k)}_L+Q^{(k)},
$$
where $\tilde H^{(k)}_L=H^{(k)}_L-(\textnormal{boundary terms})$. (Recall the boundary terms $\frac{1}{2}P_{1,2}+\frac{1}{2}P_{L-1,L}$ in the proof of Theorem \ref{thm:fs}.) There are a total of $k$ boundary interactions missing at every boundary edge. All that we need is that $0\leq \tilde H^{(k)}_L\leq H^{(k)}_L$, since all interactions $P_e$ are $\geq 0$.

 On the other hand, by the spectral theorem, translation-invariance and frustration-freeness, $A^{(k)}\geq 2k\gam_{\mathrm{loc}} \tilde H^{(k)}_L$. Hence,
$$
(H^{(k)})^2\geq  H^{(k)}+Q^{(k)}\geq H^{(k)}+2k(\gam_{\mathrm{loc}}-1) \tilde H^{(k)}_L.
$$
If $\gam_{\mathrm{loc}}\geq 1$, this implies $(H^{(k)})^2\geq H^{(k)}$ and so $\gam^{(k)}_L\geq 1$ as claimed. If $\gam_{\mathrm{loc}}\leq 1$, then we can use $\tilde H^{(k)}_L\leq H^{(k)}_L$ to obtain $(H^{(k)})^2\geq 2k\l(\gam_{\mathrm{loc}}-1+\frac{1}{2k}\r) H^{(k)}_L.$
This implies the claimed gap inequality and finishes the proof of Theorem \ref{thm:fstree}
\e{proof}

\begin{proof}[Proof of Theorem \ref{thm:maintree}]
Given Theorem \ref{thm:fstree}, it remains to show that $\gam_{\mathrm{loc}}>1-\frac{1}{2k}$ occurs with positive probability. As in the proof of Theorem \ref{thm:mainfs}, we can compute $H_{\mathrm{loc}}^2$ and apply Lemma 6.3 (ii) in \cite{FNW} to reduce this to showing
$$
\|P_{1,2}P_{2,3}-P_{1,2}\wedge P_{2,3}\|<\frac{1}{2k}.
$$
Following the proof of Theorem \ref{thm:Prandom} until the last step, we see that for every positive $0<\eps<1/4$, there is a positive probability that we have
$$
\|P_{1,2}P_{2,3}-P_{1,2}\wedge P_{2,3}\|<4r\eps.
$$
We can thus conclude by choosing $0<\eps<\frac{1}{8rk}$, say $\eps=\frac{1}{9rk}$. This proves Theorem \ref{thm:maintree}.
\e{proof}

\textit{Conclusions.}
We showed that translation-invariant spin chains with random local interactions of sufficiently small rank have a positive probability of being gapped. This observation complements the result by Movassagh \cite{M} that spatially random systems are generically gapless. It indicates that spectral gaps are more common for translation-invariant systems than for spatially random systems, at least in one dimension and on trees.

The key insight in \cite{M} is that rare local deviations can isolate arbitrarily small excitations and thereby close the gap. The mechanism used here to derive the gap is related but different: There is a positive probability that, locally, the system is extremely gap-friendly (meaning that $\|P_{1,2}P_{2,3}-P_{1,2}\wedge P_{2,3}\|<\frac{1}{2}$). Then, by the finite-size criterion, the good local behavior yields a gap for the whole system. Notice that translation-invariance enters in a fundamental way in the finite-size criterion: It suffices to check that the system is locally gap-friendly only once, because the system looks the same all over the chain. Of course, the local nature of the method is limiting: Control on the local behavior on $3$ sites is sufficient, but far from necessary, for having a spectral gap. 

To summarize, finite-size criteria are an informative, but coarse method for deriving spectral gaps. This point can also be observed in higher dimensions \cite{LN}.

It is an open problem to extend the results presented here to higher-rank interactions (presumably subject to the restriction that $r\leq \max\{d-1,d^2/4\}$ to ensure frustration-freeness), or to higher dimensions.

\section*{Acknowledgments}
We thank David Gosset, Bruno Nachtergaele, and especially Ramis Movassagh for helpful comments on a preprint version of this manuscript.


\end{document}